\documentclass[11pt]{article} 
\usepackage[pctex32]{graphics}
\usepackage{graphicx}
\usepackage{amsmath,amsxtra,amssymb,latexsym,amscd}
\usepackage[mathscr]{eucal}
\def\disp{\displaystyle}

\def\vs{\vskip0.2cm}
\def\vsm{\vskip0.15cm}
\def\vss{\vskip0.4cm}
\def\bce{\begin{center}}
\def\ece{\end{center}}

\def\beq{\begin{equation}}
\def\eeq{\end{equation}}
\def\bea{\begin{eqnarray}}
\def\eea{\end{eqnarray}}
\def\n{\noindent}
\def\nn{\nonumber}
\def\ce{\centerline}

\advance\voffset-0.2cm
\font\tit=cmb10 scaled \magstep2

\begin{document}
\parindent=0.9cm
\setlength{\baselineskip}{14truept}
\renewcommand\refname{\normalsize \centerline{REFERENCES}}
\parindent=1.05cm
\setcounter{page}{1}
\makeatletter	   
\renewcommand{\ps@plain}{%
     \renewcommand{\@oddhead}{ \thepage}
    \renewcommand{\@evenhead}{\@oddhead}%
    \renewcommand{\@oddfoot}{}
    \renewcommand{\@evenfoot}{\@oddfoot}}
    \makeatother     
\title{}
\date{}
\maketitle
\pagestyle{plain}
\pagestyle{myheadings}
\markboth{\footnotesize\it Nguyen Van Hieu and Nguyen Bich Ha}{\footnotesize \it Quantum Theory of Electron Transport Through Single-Level Quantum Dot}

$\quad$

\bce{
\vskip-4cm
\tit  QUANTUM THEORY OF ELECTRON TRANSPORT THROUGH SINGLE-LEVEL QUANTUM DOT}\ece
\vs

\ce{\bf Nguyen Van Hieu\,$^{a, b)}$ and Nguyen Bich Ha\,$^{a)}$}
\vsm
\ce{\it $^{a)}$\,College of Technology, Vietnam National University, Hanoi}
\vsm
\ce{\it $^{b)}$\,Institute of Materials Science and Institute of Physics and Electronics, VAST}

\vs
\ce{\bf Submitted to Advances in Natural Sciences}
\vsm

\vss
\leftskip1.05cm
{\small
\n {\bf Abstract.} 
A new approach in the quantum theory of few-electron nanoelectronic devices -- the S-matrix approach -- is presented in a simple example: a single-electron transistor consisting of a single-level quantum dot connected with two metallic leads through the corresponding potential barriers. The electron transport through the quantum dot due to the electron tunneling between the dot and the leads is studied. The strong Coulomb repulsion between the electrons in the dot is taken into account exactly, while the tunneling between the dot and the leads, considered as a small perturbation, is studied by means of the perturbation theory. For summing up the infinite perturbation theory series we apply the Green function technique and the Heisenberg equation of motion of the electron annihilation and creation operators. The matrix elements of the transition processes include both the direct and crossing terms, so that there is no need to use the non-crossing approximation (NCA). The explicit expression of the electron transport current is derived.

}
\leftskip0cm

\vss
\ce{\bf I. INTRODUCTION}

\vs
The transport of electrons through the single-electron transistor (SET) 
consisting of a quantum dot and two (metallic or ferromagnetic) leads under 
some bias voltage between two leads was studied in many theoretical and 
experimental works. Following Meir, Wingreen and Lee [1]
for the study of the electron transport many authors 
[2\,-\,11] used the Landauer formula [12] expressing the 
conductance in terms of the Green function of the electron in the dot, while 
Izumida \textit{et al.} [13\,-\,15] applied the Kubo linear response theory\;[16] to 
express the electrical current (or the conductivity tensor) in terms of the 
Green functions. The calculation of the Green functions has been done by 
means of different approximate methods: the truncation of the system of 
equations for many-point Green functions derived from the Heisenberg 
equation of motion\;[1,\,3], the perturbation theory with respect to the 
strong Coulomb repulsion potential\;${[2, 8, 9]}$, the non-crossing 
approximation (NCA)\;${[7]}$, the numerical renormalization group method 
\;${[10, 13, 15]}$ and the Quantum Monte Carlo technique\;${[14]}$.

In a recent work\;[17] for the study 
of the electron transport through a single-level quantum dot one of the authors has proposed another method  
based on the use of the S-matrix in the perturbation theory with respect to 
the Hamiltonian of the tunneling between the dot and the leads considered as 
the interaction Hamiltonian $H_{int}$, the strong Coulomb repulsion potential 
between two electrons in the dot being included into the part $H_{0}$ with 
exactly determined eigenvalues and eigenvectors of the total Hamiltonian 
\begin{equation}
\label{eq1}
\quad H = H_0 + H_{int}\,.
\end{equation}

As in earlier works [1\,-\,11,\,13\,-\,15] we consider the single-electron 
transistor consisting of a single-level quantum dot connected with two 
metallic leads ``a'' and ``b'' through two potential barriers. The dot and 
the leads are assumed to be kept in the quasi-equilibrium states at a fixed 
temperature and with given chemical potentials being under control. We use 
the unit system with $\hbar = c = 1\,.$

\def\sat{\vskip-0.5cm}

Denote $c_\sigma $ and $c_\sigma ^ + $ the annihilation and creation 
operators of the electron with the spin projection $\sigma = \uparrow , 
\downarrow $ at the single energy level $E$ in the quantum dot, $a_\sigma 
({\pmb k})$, $b_\sigma ({\pmb k})$ and $a_\sigma ^ + ({\pmb k})$, $b_\sigma 
^ + ({\pmb k})$ those of electrons with the spin projection \textit{$\sigma $} and the 
momentum \textbf{k} in the leads ``a'' and ``b'', resp. Following earlier 
works ${[1 - 11, 13 - 15]}$ we assume that
\begin{equation}
H_0 = E\sum\limits_\sigma {c_\sigma ^ + c_\sigma } + U{\kern 1pt} n_ 
\uparrow {\kern 1pt} n_ \downarrow \; + \sum\limits_{\pmb k} 
{\sum\limits_\sigma {\left\{ {\;E_a ({\pmb k})\,a_\sigma ^ + 
({\pmb k})\,a_\sigma ({\pmb k}) + E_b ({\pmb k})\,b_\sigma ^ + 
({\pmb k})\,b_\sigma ({\pmb k})} \right\},} } 
\end{equation}
\begin{equation}
\;n_\sigma = c_\sigma ^ + \,c_\sigma ,
\end{equation}
\begin{equation}
H_{int} = \sum\limits_{\pmb k} {\sum\limits_\sigma {\left\{ {\;V_a 
({\pmb k})\,a_\sigma ^ + ({\pmb k})\,c_\sigma + V_a ({\pmb k})^ * \,c_\sigma 
^ + \,a_\sigma ({\pmb k})} \right.} } \; + V_b ({\pmb k})\,b_\sigma ^ + 
({\pmb k})\,c_\sigma + V_b ({\pmb k})^ * \,c_\sigma ^ + \,b_\sigma \left. 
{({\pmb k})} \right\}.
\end{equation}

In the previous report ${[17]}$ the electron current from one lead to 
another one in the presence of some bias voltage between two leads was 
calculated in the second order of the perturbation theory with respect to 
the interaction Hamiltonian (4). In this work the sums of the series of high 
order terms of the perturbation theory in the ladder approximation are 
calculated by means of the Green function technique with the use of the 
Heisenberg equation of motion for the electron annihilation and creation 
operators. As we shall see, the matrix elements of the S-matrix contain both 
the ``direct'' and ``crossing'' terms, so that in the framework of the 
present S-matrix approach there is no need to assume the non-crossing 
approximation. The ``crossing'' terms 
describe the co-tunneling processes. In many cases their contributions are significant.

\vss
\ce{\bf II. PERTUBATION THEORY IN THE SECOND ORDER}

\vs
In order to apply the perturbation theory we introduce the electron 
annihilation and creation operators as well as the interaction Hamiltonian 
in the interaction representation
\[
c_\sigma ^ + (t) = e^{iH_0 t}\,c_\sigma ^ + \,e^{ - iH_0 t},\quad
c_\sigma (t) = e^{iH_0 t}c_\sigma e^{ - iH_0 t},
\]
\[
a_\sigma ^ + ({\pmb k},t) = e^{iH_0 t}\,a_\sigma ^ + ({\pmb k})\,e^{ - iH_0 
t},\quad
a_\sigma ({\pmb k},t) = e^{iH_0 t}a_\sigma ({\pmb k})\,e^{ - iH_0 t},
\]
\begin{equation}
\label{eq2}
b_\sigma ^ + ({\pmb k},t) = e^{iH_0 t}\,b_\sigma ^ + ({\pmb k})\,e^{ - iH_0 
t},\quad
b_\sigma ({\pmb k},t) = e^{iH_0 t}b_\sigma ({\pmb k})\,e^{ - iH_0 t},
\end{equation}
\[
H_{int} (t) = e^{iH_0 t}H_{int} e^{ - iH_0 t}.
\]

\n The S-matrix in the perturbation theory with respect to the 
interaction Hamiltonian (4) equals
\begin{equation}
S = T\bigg\{ {\exp \bigg( { - i\int\limits_{ - \infty }^\infty {H_{int} 
(t){\kern 1pt} dt} } \bigg)} \bigg\} = \sum\limits_{n = 0}^\infty {\frac{( 
- i)^n}{n\,!}} \int\limits_{ - \infty }^\infty {dt_1 ...} \int\limits_{ - 
\infty }^\infty {dt_n } \;T\left\{ {H_{int} (t_1 )...H_{int} (t_n )} 
\right\}.
\end{equation}

\n We set
\begin{equation}
S = 1 - iR ,
\end{equation}
and write
\begin{equation}
R = \sum\limits_{i = 1}^\infty {R^{(n)}},
\end{equation}
where $R^{(n)}$ denotes the contribution of all \textit{n-th }order terms. In the first 
order of the perturbation theory we have
\begin{eqnarray}
R^{(1)} = &\sum\limits_\sigma {\sum\limits_{\pmb k} {\int\limits_{ - \infty 
}^\infty {dt\;\left\{ {\;\,V_a ({\pmb k})\,a_\sigma ^ + 
({\pmb k},t)\,c_\sigma (t) + V_a ({\pmb k})^ * c_\sigma ^ + (t)\,a_\sigma 
({\pmb k},t)} \right.} } } + \\ 
&  \quad \quad \quad \quad\;\; + V_b ({\pmb k})\,b_\sigma ^ + 
({\pmb k},t)\,c_\sigma (t) + V_b ({\pmb k})^ * c_\sigma ^ + (t)\,b_\sigma 
\left. {({\pmb k},t)} \right\},\nn 
\end{eqnarray}
while in the second order
\sat
\begin{eqnarray}
&& R^{(2)} = - \;\,\frac{i}{2}\sum\limits_{\sigma _1 } {\sum\limits_{\sigma _2 
} {\sum\limits_{{\pmb k}_1 } {\sum\limits_{{\pmb k}_2 } {\int\limits_{ - 
\infty }^\infty {dt_1 \int\limits_{ - \infty }^\infty {dt_2 } } } } } } \;\nn 
\\ 
 &&\quad \quad \; \quad \quad T\,\Big\{ {\big[ {\;V_a ({\pmb k}_1 )\,} 
} a_{\sigma _1 }^ + ({\pmb k}_1 ,t_1 )\,c_{\sigma _1 } (t_1 ) 
+ V_a ({\pmb k}_1 )^ * c_{\sigma _1 }^ + (t_1 )\,a_{\sigma _1 } ({\pmb k}_1 
,t_1 ) \nn\\ 
&& \quad \quad \quad \quad \quad \quad + V_b ({\pmb k}_1 )\,b_{\sigma _1 }^ + 
({\pmb k}_1 ,t_1 )\,c_{\sigma _1 } (t_1 ) + V_b ({\pmb k}_1 )^ * c_{\sigma 
_1 }^ + (t_1 )\,b_{\sigma _1 }  {({\pmb k}_1 ,t_1 )} \big ] \\ 
&& \quad \quad \quad \quad \quad \quad \;\;\big[ {\;V_a ({\pmb k}_2 )} 
\,a_{\sigma _2 }^ + ({\pmb k}_2 ,t_2 )\,c_{\sigma _2 } (t_2 ) + V_a 
({\pmb k}_2 )^ * c_{\sigma _2 }^ + (t_2 )\,a_{\sigma _2 } ({\pmb k}_2 ,t_2 ) \nn
\\ 
&& \quad \quad \quad \quad \quad \quad + V_b ({\pmb k}_2 )\,b_{\sigma _2 }^ + 
({\pmb k}_2 ,t_2 )\,c_{\sigma _2 } (t_2 ) + V_b ({\pmb k}_2 )^ * c_{\sigma 
_2 }^ + (t_2 )\, { {b_{\sigma _2 } ({\pmb k}_2 ,t_2 )} 
\big]} \Big\}.\nn 
 \end{eqnarray}

If the initial and final states $\left| i \right\rangle $ and $\left| f 
\right\rangle $ in some matrix element $\left\langle {f\,\left| {\,R\,} 
\right|\,i} \right\rangle $ of the scattering operator $R$ contain one and the 
same subsystem of $n$ particles $a_\sigma ({\pmb k})$ and $m$ particles $b_\sigma 
({\pmb k})$, ($n, m$=0,1,2,\ldots ) which do not participate in the interaction 
process, then this matrix element can be written in the form
\begin{align}
& \big\langle f |\,R\,| i \big\rangle = \sum\limits_P {( - 
1)^P\big\langle {a_{{\sigma }'_1 } ({{\pmb k}'}_1 )\,\left| {\,a_{\sigma _1 
} ({\pmb k}_1 )} \right.} \big\rangle } \cdot \cdot \cdot \big\langle 
{a_{{\sigma }'_n } ({{\pmb k}'}_n )\,\left| {\,a_{\sigma _n } ({\pmb k}_n )} 
\right.} \big\rangle \times\\ 
& \qquad \quad \quad\; \times \big\langle {b_{{\sigma }'_{n + 1} } 
({{\pmb k}'}_{n + 1} )\,\left| {\,b_{\sigma _{n + 1} } ({\pmb k}_{n + 1} )} 
\right.} \big\rangle \cdot \cdot \cdot \big\langle {b_{{\sigma }'_{n 
+ m} } ({{\pmb k}'}_{n + m} )\,\left| {\,b_{\sigma _{n + m} } ({\pmb k}_{n + m} 
)} \right.} \big\rangle \,\big\langle {\tilde {f}\left| {\,R\,} 
\right|\tilde {i}} \big\rangle \,,\nn 
\end{align}
where $\left\{\,{\tilde {i}\,} \right\}$ and $\left\{\,{\tilde {f}\,} 
\right\}$ are the subsystems complementary to those of the non-interacting 
particles $a_{\sigma _1 } ({\pmb k}_1 ),\;...\;a_{\sigma _n } ({\pmb k}_n 
),\;b_{\sigma _{n + 1} } ({\pmb k}_{n + 1} ),\;...\;b_{\sigma _{n + m} } 
({\pmb k}_{n + m} )$ in the initial and final states, resp., $P$ denotes the 
antisymmetrization with respect to the identical particles of each kind 
$a_\sigma ({\pmb k})$ and $b_\sigma ({\pmb k})$. The matrix elements in this 
factorized and antisymmetrized form are called reducible ones contrary to 
the irreducible matrix elements which cannot be expressed in the 
above--mentioned form (11). In the study of the observable physical 
quantities of the system it is necessary and sufficient to derive the 
explicit expressions of the irreducible matrix elements of the scattering 
operator $R$.

Consider now in detail the transport of the electron through the quantum dot 
-- the transition of the electron from one lead into another one ($a_\sigma 
({\pmb k}) \to b_{\sigma'} ({{\pmb k}'})$, for example) via the 
intermediate states of the electron in the quantum dot (virtual particles 
$c_{\sigma }$) by means of the tunneling of the 
electrons between the leads and the dot. In the second order of the 
perturbation theory we have following irreducible matrix elements:
\begin{equation}
\begin{array}{l}
\left\langle {b_{\sigma'} ({\pmb{k}'})\left| {R^{(2)}} \right|a_\sigma 
({\pmb k})} \right\rangle = \\
\quad - \;iV_a ({\pmb k})^ * V_b 
({{\pmb k}'})\int\limits_{ - \infty }^\infty {dt_1 } \int\limits_{ - \infty 
}^\infty {dt_2 \,e^{i[E_b ({{\pmb k}'})\,t_1 - E_a ({\pmb k})\,t_2 
]}\left\langle {0\,\left| {T\{c_{\sigma'} (t_1 )\,c_\sigma ^ + (t_2 
)\}\left| 0 \right.} \right.} \right\rangle },
\end{array}
\end{equation}
\begin{equation}
\label{eq3}
\begin{array}{l}
 \;\left\langle {c_{{\sigma'}_2 } b_{{\sigma }'_1 } ({\pmb{k}'})\left| 
{R^{(2)}} \right|a_{\sigma _1 } ({\pmb k})\,c_{\sigma _2 } } \right\rangle = 
\\ 
 \quad - \;i\;V_a ({\pmb k})^ * \,V_b ({{\pmb k}'})\int\limits_{ - \infty 
}^\infty {dt_1 } \int\limits_{ - \infty }^\infty {dt_2 \,e^{i[E_b 
({{\pmb k}'})\,t_1 - E_a ({\pmb k})\,t_2 ]}\left\langle {c_{{\sigma }'_2 } 
\left| {T\{c_{{\sigma }'_1 } (t_1 )\,c_{\sigma _1 }^ + (t_2 )\}\left| 
{c_{\sigma _2 } } \right.} \right.} \right\rangle }, 
 \end{array}
\end{equation}
\begin{equation}
\begin{array}{l}
 \left\langle {c_ \downarrow c_ \uparrow b_{\sigma'} ({{\pmb k}'})\left| 
{R^{(2)}} \right|a_\sigma ({\pmb k})\,c_ \uparrow c_ \downarrow } 
\right\rangle = \\ 
 \quad - \;i\;V_a ({\pmb k})^ * \,V_b ({{\pmb k}'})\int\limits_{ - \infty 
}^\infty {dt_1 } \int\limits_{ - \infty }^\infty {dt_2 \,e^{i[E_b 
({{\pmb k}'})\,t_1 - E_a ({\pmb k})\,t_2 ]}\left\langle {c_ \downarrow c_ 
\uparrow \left| {T\{c_{\sigma'} (t_1 )\,c_\sigma ^ + (t_2 )\}\left| {c_ 
\uparrow c_ \downarrow } \right.} \right.} \right\rangle } . \\ 
 \end{array}
\end{equation}

\n The problem is now reduced to the calculation of the matrix elements of the 
chronological product of two operators $c_{\sigma'} ({t}')$ and $c_\sigma ^ 
+ (t)$ between different electronic states of the quantum dot: the empty 
state $\left| 0 \right\rangle $, the singly occupied states $\left| 
{c_\sigma } \right\rangle $ and the doubly occupied one $\left| {c_ \uparrow 
c_ \downarrow } \right\rangle $.

The matrix element
\begin{equation}
G_{{\sigma }'\sigma }^{(0)} ({t}' - t) = - i\left\langle {0\;\left| 
{T\{c_{\sigma'} ({t}')\,c_\sigma ^ + (t)\}} \right|\;0} \right\rangle 
\end{equation}
is the usual Green function of the electron in the quantum dot, while two other 
matrix elements
\begin{equation}
G_{{\sigma }'_2 {\sigma }'_1 \sigma _1 \sigma _2 }^{(\ref{eq1})} ({t}' - t) = - 
i\left\langle {c_{{\sigma }'_2 } \left| {T\{c_{{\sigma }'_1 } 
({t}')\,c_{\sigma _1 }^ + (t)\}} \right|c_{\sigma _2 } } \right\rangle 
\end{equation}
and
\begin{equation}
G_{{\sigma }'\sigma }^{(2)} ({t}' - t) = - i\left\langle {c_ \downarrow c_ 
\uparrow \left| {T\{c_{\sigma'} ({t}')\,c_\sigma ^ + (t)\}} \right|c_ 
\uparrow c_ \downarrow } \right\rangle 
\end{equation}
are only similar to the usual Green function and might be called the generalized 
Green functions. They can be exactly calculated without any approximation 
with respect to the large Coulomb repulsion potential $U$.

The chronological product of two operators $c_{\sigma'} ({t}')$ 
and $c_\sigma ^ + (t)$ has the explicit form
\begin{equation}
T\{c_{\sigma'} ({t}')\,c_\sigma ^ + (t)\} = \theta ({t}' - t)\,c_{\sigma'} 
({t}')\,c_\sigma ^ + (t) - \theta (t - {t}')\,c_\sigma ^ + (t)\,c_{\sigma'} 
({t}')\;.\end{equation}

\n The first term in the r. h. s. of the relation (18) describes the creation 
of an electron in the quantum dot at the moment $t$ (due to the tunneling of 
this electron from the lead ``$a$") and then the subsequent annihilation of 
this electron $(\sigma =\sigma^\prime)$ or another electron $(\sigma \not=\sigma^\prime)$ in the quantum dot at a later moment $t' > t$ (due to its tunneling 
to the lead ``$b$"), while the second term describes the annihilation of an 
electron in the quantum dot at the moment $t'$ (due to its tunneling to the lead 
``$b$") and the subsequent creation of another electron in the quantum dot at 
a later moment $t' > t$ (due to the tunneling of this electron from the lead 
``$a$"). The first term is called the direct one, while the second one is 
called the crossing term. They are represented 
by the diagrams in Fig.\;a and Fig.\;1b. 

\vs
\centerline{\includegraphics[width=14.5cm,height=5cm]{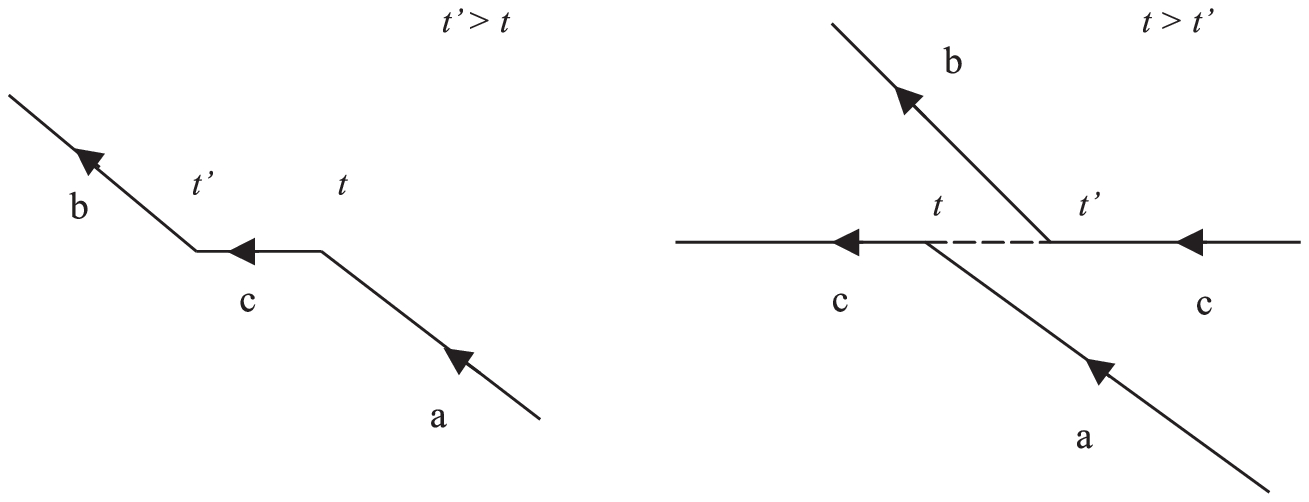}}

\vs
\ce{\small\textbf{Fig.\,1a} \hskip4.6cm \textbf{Fig.\,1b}}
\vskip0.3cm

The ability to take into account the contributions of the crossing terms is 
an advantage of the present S-matrix method. If the initial state is the 
empty one $\left| 0 \right\rangle $, then there is no electron in the 
quantum dot to be annihilated and the crossing term does not appear: 
\begin{equation}
\label{eq4}
\left\langle {0\left| {T\{c_{\sigma'} ({t}')\,c_\sigma ^ + (t)\}} \right|0} 
\right\rangle = \theta ({t}' - t)\,e^{ - iE({t}' - t)}\delta _{\sigma 
{\sigma }'} \,.
\end{equation}

\n On the contrary, if the initial state is the doubly occupied one $\left| {c_ 
\uparrow c_ \downarrow } \right\rangle $, then it is impossible to add one 
more electron into the quantum dot and only the crossing term gives the 
contribution: 
\begin{equation}
\label{eq5}
\left\langle {c_ \downarrow c_ \uparrow \left| {T\{c_{\sigma'} 
({t}')\,c_\sigma ^ + (t)\}} \right|c_ \uparrow c_ \downarrow } \right\rangle 
= - \,\theta (t - {t}')\,e^{ - i(E + U)\,({t}' - t)}\delta _{\sigma {\sigma 
}'} \,.
\end{equation}

\n In the case of the singly occupied initial states $\left| {c_{\sigma _2 } } 
\right\rangle $ both the direct and crossing terms can give non-vanishing 
contributions depending on the spin configuration of the electron system: 
\begin{equation}
\begin{array}{l}
 \left\langle {c_{{\sigma }'_2 } \left| {T\{c_{{\sigma }'_1 } 
({t}')\,c_{\sigma _1 }^ + (t)\}} \right|c_{\sigma _2 } } \right\rangle = 
\quad \theta ({t}' - t)\,e^{ - i(E + U)\,({t}' - t)}\delta _{ - \sigma _1 
\sigma _2 } \delta _{ - {\sigma }'_1 {\sigma }'_2 } \left[ {\delta _{{\sigma 
}'_1 \sigma _1 } - \delta _{{\sigma }'_1 \sigma _2 } } \right] \\ 
 \quad \quad \quad \quad \quad \quad \quad \quad \quad \quad \quad \; - 
\theta (t - {t}')\,e^{ - iE({t}' - t)}\delta _{{\sigma }'_1 \sigma _2 } 
\delta _{{\sigma }'_2 \sigma _1}\,.  
 \end{array}
\end{equation}

Introduce the Fourier transformations of the usual and generalized Green 
functions
\begin{equation}
G_{{\sigma }'\sigma }^{(0)} ({t}' - t) = \frac{1}{2\pi }\int {d\omega \,e^{ 
- i\omega ({t}' - t)}\tilde {G}_{{\sigma }'\sigma }^{(0)} } (\omega 
),
\end{equation}
\begin{equation}
G_{{\sigma }'_2 {\sigma }'_1 \sigma _1 \sigma _2 }^{(\ref{eq1})} ({t}' - t) = 
\frac{1}{2\pi }\int {d\omega \,e^{ - i\omega ({t}' - t)}} \tilde 
{G}_{{\sigma }'_2 {\sigma }'_1 \sigma _1 \sigma _2 }^{(\ref{eq1})} (\omega ),
\end{equation}
\begin{equation}
G_{{\sigma }'\sigma }^{(2)} ({t}' - t) = \frac{1}{2\pi }\int {d\omega \,e^{ 
- i\omega (t_1 - t_2 )}\tilde {G}_{{\sigma }'\sigma }^{(2)} (\omega ).} 
\end{equation}

\n From the expressions (\ref{eq4})\,-\,(21) of the Green functions it follows that 
their Fourier transforms equal
\begin{equation}
\tilde {G}_{{\sigma }'\sigma }^{(0)} (\omega ) = \frac{\delta _{{\sigma 
}'\sigma } }{\omega - E + io}\ ,
\end{equation}
\vskip-0.5cm
\begin{equation}
\begin{array}{l}
 \\ 
 \tilde {G}_{{\sigma }'_2 {\sigma }'_1 \sigma _1 \sigma _2 }^{(\ref{eq1})} (\omega ) 
= \dfrac{\delta _{ - \;\sigma _1 \sigma _2 } \delta _{ - \;{\sigma }'_1 
{\sigma }'_2 } [ {\delta _{{\sigma }'_1 \sigma _1 } - \delta _{{\sigma 
}'_1 \sigma _2 } } ]}{\omega - E - U + io} + \dfrac{\delta _{{\sigma 
}'_1 \sigma _2 } \delta _{{\sigma }'_2 \sigma _1 } }{\omega - E - io}\ , 
 \end{array}
\end{equation}
\begin{equation}
\tilde {G}_{{\sigma }'\sigma }^{(2)} (\omega ) = \frac{\delta _{{\sigma 
}'\sigma } }{\omega - E - U - io}\ \cdot 
\end{equation}

It is easy to express the matrix elements (13)\,-\,(14) in terms of these 
Fourier transforms. Finally we obtain following relevant non-vanishing 
irreducible matrix elements contributing to the transitions of a particle 
$a_\sigma ({\pmb k})$ into a particle $b_{\sigma '} ({{\pmb k}'})$ via the 
intermediate virtual particle $c_{{\sigma ''}} $ in the second order of the 
perturbation theory: 
\begin{equation}
\begin{array}{l}
 \left\langle {b_ \uparrow ({{\pmb k}'})\left| {R^{(2)}} \right|a_ \uparrow 
({\pmb k})} \right\rangle = \left\langle {b_ \downarrow ({{\pmb k}'})\left| 
{R^{(2)}} \right|a_ \downarrow ({\pmb k})} \right\rangle = \\ 
 \quad \quad \quad \quad \quad \quad = 2\pi \;\delta [E_a ({\pmb k}) - 
E_b ({{\pmb k}'})]\;V_a ({\pmb k})^ * V_b ({{\pmb k}'}) \cdot \dfrac{1}{E_a 
({\pmb k}) - E + io}\ ,  
 \end{array}
\end{equation}
\vskip-0.7cm
\begin{equation}
\begin{array}{l}
 \left\langle {c_ \downarrow b_ \uparrow ({{\pmb k}'})\left| {R^{(2)}} 
\right|a_ \uparrow ({\pmb k})\,c_ \downarrow } \right\rangle = \left\langle 
{c_ \uparrow b_ \downarrow ({{\pmb k}'})\left| {R^{(2)}} \right|a_ 
\downarrow ({\pmb k})\,c_ \uparrow } \right\rangle = \\ 
 \quad \quad \quad \quad \quad = 2\pi \;\delta [E_a ({\pmb k}) - E_b 
({{\pmb k}'})]\;V_a ({\pmb k})^ * V_b ({{\pmb k}'}) \cdot \dfrac{1}{E_a 
({\pmb k}) - E - U + io}\ ,  
 \end{array}
\end{equation}
\vskip-0.7cm
\begin{equation}
\begin{array}{l}
 \left\langle {c_ \uparrow b_ \uparrow ({{\pmb k}'})\left| {R^{(2)}} 
\right|a_ \uparrow ({\pmb k})\,c_ \uparrow } \right\rangle = \left\langle 
{c_ \downarrow b_ \downarrow ({{\pmb k}'})\left| {R^{(2)}} \right|a_ 
\downarrow ({\pmb k})\,c_ \downarrow } \right\rangle = \\ 
 \quad \quad \quad \quad \quad = 2\pi \;\delta [E_a ({\pmb k}) - E_b 
({{\pmb k}'})]\;V_a ({\pmb k})^ * V_b ({{\pmb k}'}) \cdot \dfrac{1}{E_a 
({\pmb k}) - E - io}\ ,  
 \end{array}
\end{equation}
\vskip-0.7cm
\begin{equation}
\label{eq6}
\begin{array}{l}
 \left\langle {c_ \uparrow b_ \downarrow ({{\pmb k}'})\left| {R^{(2)}} 
\right|a_ \uparrow ({\pmb k})\,c_ \downarrow } \right\rangle = \left\langle 
{c_ \downarrow b_ \uparrow ({{\pmb k}'})\left| {R^{(2)}} \right|a_ 
\downarrow ({\pmb k})\,c_ \uparrow } \right\rangle = \\ 
 \;\,  = 2\pi \;\delta \,[E_a ({\pmb k}) - E_b 
({{\pmb k}'})]\;V_a ({\pmb k})^ * V_b ({{\pmb k}'})\left[ { - \;\dfrac{1}{E_a 
({\pmb k}) - E - U + io} + \dfrac{1}{E_a ({\pmb k}) - E - io}} \right]\;,  
 \end{array}
\end{equation}
\vskip-0.8cm
\begin{equation}
\begin{array}{l}
 \left\langle {c_ \downarrow c_ \uparrow b_ \uparrow ({{\pmb k}'})\left| 
{R^{(2)}} \right|a_ \uparrow ({\pmb k})\,c_ \uparrow c_ \downarrow } 
\right\rangle = \left\langle {c_ \downarrow c_ \uparrow b_ \downarrow 
({{\pmb k}'})\left| {R^{(2)}} \right|a_ \downarrow ({\pmb k})\,c_ \uparrow 
c_ \downarrow } \right\rangle = \\ 
 \quad \quad \quad \quad \quad \quad = 2\pi \;\delta [E_a ({\pmb k}) - E_b 
({{\pmb k}'})]\;V_a ({\pmb k})^ * V_b ({{\pmb k}'}) \cdot \dfrac{1}{E_a 
({\pmb k}) - E - U - io}\ \cdot  
 \end{array}
\end{equation}
Note that the matrix element (\ref{eq6}) describes the spin-flip scattering process 
inside the dot while other matrix elements are those of the non-spin-flip ones.

Using the expressions (28)\,-\,(32) of the matrix elements of the scattering 
operator it is easy to calculate the transition rate. If the matrix element 
of the transition from an initial pure state $\left| i \right\rangle $ to a 
final one $\left| f \right\rangle $ has the form
\begin{equation}
\left\langle {f\,\left| {\,R\,} \right|\,i} \right\rangle = 2\pi \,\delta 
[E_i - E_f ]\;M_{i \to f} \ ,
\end{equation}
then the transition rate in an unit time interval equals
\begin{equation*}
W_{i\, \to f} = 2\pi \,\delta [E_i - E_f ]\;\left| {M_{i\, \to f} } 
\right|^2\,.
\end{equation*}
If the 
initial state \pagebreak is a mixed equilibrium state $I$ containing the pure state 
$\left| i \right\rangle $ with a weight $w_{i}$\,, then the transition rate 
from this mixed state $I$ to the final states $\left| f \right\rangle $ 
belonging to some class $F$ equals
\begin{equation}
W_{I\, \to F} = \sum\limits_{i \in I} {\sum\limits_{f \in F} 2 } \pi 
\,\delta [E_i - E_f ]\;\left| {M_{i\, \to f} } \right|^2\,w_i \;.
\end{equation}

Let us apply this formula to the transitions between the initial and final 
states presented in the matrix elements (28)\,-\,(32). The statistical 
weights of these pure states in the initial mixed equilibrium state are 
given in the Table I with 
\begin{equation}
\begin{array}{l}
 Z\,\; = Z_a \,Z_b \,Z_c \quad ,\par \\ 
 Z_a = [1 + e^{ - \beta \,{E}'_a (\pmb{k})}]^2\, , \\ 
 Z_b = [1 + e^{ - \beta \,{E}'_b (\pmb{k})}]^2\, , \\ 
 Z_c = 1 + 2e^{ - \beta \,{E}'} + e^{ - \beta (2{E}' + U)}\, , \\ 
 \end{array}
\end{equation}

{\bf Table I}
\begin{center}
\begin{tabular}
{|p{159pt}|p{182pt}|}
\hline\par
&\\
State& 
Weight \\
&\\
\hline \par
\raisebox{-2ex}{$\left| {a_\sigma ({\pmb k})} \right\rangle $}&\par 
\raisebox{-2ex}{${e^{ - \beta {E}'_a ({\pmb k})}}/{Z}$}\par \\
\hline \par
\raisebox{-2ex}{$\left| {a_ \uparrow ({\pmb k})\,a_ \downarrow ({\pmb k})} \right\rangle $}&
\raisebox{-2ex}{${e^{-\beta 2 E^{\prime}_a({\pmb k})}}/{Z}$}\par \\
\hline \par
\raisebox{-2ex}{$\left| {a_\sigma ({\pmb k})\,b_{\sigma '} ({\pmb k}')} \right\rangle $}& 
\raisebox{-2ex}{${e^{-\beta[E^{\prime}_a(\pmb k)+ E^{\prime}_b(\pmb k^\prime)}]}/{Z}$}\par \\
\hline \par
\raisebox{-2ex}{$\left| {a_ \uparrow ({\pmb k})\,b_{\sigma'} ({{\pmb k}'})\,a_ \downarrow (\pmb{k})} \right\rangle$}& 
\raisebox{-2ex}{${e^{ - \beta [2{E}'_a (\pmb{k}) + {E}'_b (\pmb{{k}'})]}}/{Z}$}\par \\
\hline \par
\raisebox{-2ex}{$\left| {a_\sigma (\pmb{k})\,c_{{\sigma }'} } \right\rangle $}& 
\raisebox{-2ex}{${e^{ - \beta [{E}'_a (\pmb{k}) + {E}']}}/{Z}$}\par \\
\hline \par
\raisebox{-2ex}{$\left| {a_ \uparrow (\pmb{k})\,c_{{\sigma }'} \,a_ \downarrow (\pmb{k})} \right\rangle $}& 
\raisebox{-2ex}{${e^{ - \beta [2{E}'_a (\pmb{k}) + {E}']}}/{Z}$}\par \\
\hline \par
\raisebox{-2ex}{$\left| {a_\sigma (\pmb{k})\,c_{{\sigma }''} \,b_{\sigma '} (\pmb{{k}'})} \right\rangle $}&
\raisebox{-2ex}{${e^{ - \beta [{E}'_a (\pmb{k}) + {E}'_b (\pmb{{k}'}) + {E}']}}/{Z}$}\par \\
\hline \par
\raisebox{-2ex}{$\left| {a_\uparrow (\pmb{k})\,c_{{\sigma }''} \,b_{\sigma'} (\pmb{{k}'})\,a_ \downarrow (\pmb{k})} \right\rangle$}&
\raisebox{-2ex}{${e^{-\beta [2{E}'_a (\pmb{k}) + {E}'_b (\pmb{{k}'}) + {E}']}}/{Z}$}\par \\
\hline \par
\raisebox{-2ex}{$\left| {a_\sigma (\pmb{k})\,c_ \uparrow \,c_ \downarrow } \right\rangle$}&
\raisebox{-2ex}{${e^{ - \beta [{E}'_a (\pmb{k}) + 2{E}' + U]}}/{Z}$}\par \\
\hline \par
\raisebox{-2ex}{$\left| {a_ \uparrow (\pmb{k})\,c_ \uparrow \,c_ \downarrow \,a_ \downarrow (\pmb{k})} \right\rangle $}& 
\raisebox{-2ex}{${e^{ - \beta [2{E}'_a (\pmb{k}) + 2{E}' + U]}}/{Z}$}\par \\
\hline \par
\raisebox{-2ex}{$\left| {a_\sigma (\pmb{k})\,c_ \uparrow \,c_ \downarrow \,b_{\sigma'} (\pmb{{k}'})} \right\rangle $}&
\raisebox{-2ex}{${e^{ - \beta [{E}'_a (\pmb{k}) + {E}'_b (\pmb{{k}'}) + 2{E}' + U]}}/{Z}$}\par \\
\hline \par
\raisebox{-2ex}{$\left| {a_ \uparrow (\pmb{k})\,c_ \uparrow \,c_ \downarrow \,b_{\sigma '} (\pmb{{k}'})\,a_ \downarrow (\pmb{k})} \right\rangle$} &
\raisebox{-2ex}{${e^{ - \beta [2{E}'_a (\pmb{k}) + {E}'_b (\pmb{{k}'}) + 2{E}' + U]}}/{Z}$}\par \\
\hline \par
\end{tabular}
\end{center}

\n where
\begin{equation}
\begin{array}{l}
 {E}'_a (\pmb{k}) = E_a (\pmb{k}) - \mu _a \, , \\ 
 {E}'_b (\pmb{k}) = E_b (\pmb{k}) - \mu _b \, , \\ 
 \quad \;\,{E}' = E - \mu _c \, , \\ 
 \end{array}
\end{equation}
\textit{$\mu $}$_{a}$, \textit{$\mu $}$_{b}$ and \textit{$\mu $}$_{c}$ being the chemical potentials in the leads ``a", 
``b" and in the quantum dot, resp. Substituting the expressions of the 
matrix elements (28)\,-\,(32) and the statistical weights in the Table I to 
the r.\,h.\,s. of the formula (34) we obtain following expression of the 
probability of the transition of electrons from the lead ``a" to the lead 
``b" through its intermediate states in the quantum dot:
\begin{equation}
W_{a\, \to \,b} = \disp\frac{1}{\pi }\int\limits_{ - \infty }^\infty {d\omega 
\,f_a (\omega )\,[1 - f_b (\omega )]\;W(\omega )} \, ,
\end{equation}
\sat
\begin{eqnarray}
W(\omega ) &=& \sum\limits_{\pmb k} {2\pi \,\delta \,[\omega - E_a 
(\pmb{k})]\;\left| {V_a (\pmb{k})} \right|^2} \,\sum\limits_{{\pmb k}'} 
{2\pi \, \delta \,[\omega - E_b ({{\pmb k}'})]\;\left| {V_b ({{\pmb k}'})} 
\right|^2} \, . \nn \\
&&
\;\,\frac{1}{Z_c }\left\{ {\left| {\frac{1}{\omega - E + io\,}} \right|^2 
+ } \right.e^{ - \,\beta {E}'}\,\left( {\left| {\frac{1}{\omega - E - U + 
io\,}} \right|^2} \right. + \left| {\frac{1}{\omega - E - io\,}} \right|^2 + 
\\ 
 &&+ \left. {\left| {\frac{1}{\omega - E - U + io} - \frac{1}{\omega - E - 
io\;}} \right|^2} \right)\; + e^{ - \,\beta (2{E}' + U)}\,\left. {\left| 
{\frac{1}{\omega - E - U - io\;}} \right|^2} \right\} \, . \nn
\end{eqnarray}

\n Here $f_{a}$(\textit{$\omega $}) and $f_{b}$(\textit{$\omega $}) are the Fermi 
distribution functions of the electrons in the leads ``$a$`` and ``b``, resp.,
\begin{equation}
f_b (\omega ) = \frac{e^{ - \beta (\omega - \mu _b )}}{1 + e^{ - \beta 
(\omega - \mu _b )}}\;.\quad  
f_a (\omega ) = \frac{e^{ - \beta (\omega - \mu _a )}}{1 + e^{ - \beta 
(\omega - \mu _a )}}\; .
\end{equation}
For the reverse transition of electrons from the lead ``b" to the lead 
``a" we have a similar expression of the transition rate
\begin{equation}
W_{b\, \to \,a} = \frac{1}{\pi }\int\limits_{ - \infty }^\infty {d\omega 
\,f_b (\omega )\,[1 - f_a (\omega )]\;W(\omega )} \,.
\end{equation}

If $\mu_{a} > \mu_{b}$, then $W_{a \to b} > W_{b \to a}$ and the effective 
transition of electrons from the lead ``a" to the lead ``b" generates the 
electrical current
\begin{equation}
J_e = e\,[\,W_{a\, \to \,b} - W_{b\, \to \,a} \,] = e\frac{1}{\pi 
}\int\limits_{ - \infty }^\infty {d\omega \,[f_a (\omega ) - f_b (\omega 
)]\;W(\omega )}\;
\end{equation}
from the lead ``b" to the lead ``a", where $e$ is the absolute value of the 
electron charge.

In the expression (38) there are the contributions of both direct and 
crossing terms of the chronological product (21). If we neglect the crossing 
terms, then we have the non-crossing approximation (NCA). In this approximation 
$W$(\textit{$\omega $}) is replaced by the following expression
\begin{eqnarray}
\label{eq7}
W_{NCA} (\omega ) &= \sum\limits_{\pmb k} {2\pi \,\delta \,[\omega - E_a 
(\pmb{k})]\;\left| {V_a (\pmb{k})} \right|^2} \,\sum\limits_{\pmb{k}'} 
{2\pi \,\delta \,[\omega - E_b (\pmb{{k}'})]\;\left| {V_b (\pmb{{k}'})} 
\right|^2} \, \times \nn\\
&\;\;\times \frac{1}{Z_c }\left\{ {\left| {\frac{1}{\omega - E + io\,}} \right|^2 
+ } \right.2e^{ - \,\beta {E}'}\,\left. {\left| {\frac{1}{\omega - E - U + 
io\;}} \right|^2} \right\} \; \cdot 
\end{eqnarray}
Obviously, if $e^{ - \beta E'}$ is not very small, then the contribution of 
the crossing terms might be comparable to that of the direct ones.

\vss
\ce{\bf III. HIGH ORDER APPROXIMATIONS OF }
\ce{\bf THE PERTURBATION THEORY. GREEN FUNCTION TECHNIQUE}

\vs
The expression in the r.\,h.\,s. of the formula (38) exhibits the sharp 
resonances at the values of $\omega$ near the energy levels of the 
electron in the dot $\omega  \approx  E$ and $\omega  \approx  E+ U$. However, at
these values of $\omega$ the broadening of the energy levels due to the tunneling 
of the electrons from (to) the dot to (from) the leads must be taken into 
account. This can be done by summing up the contributions of the high order 
terms in the ladder approximation. The most convenient way to do that is the 
application of the Green function technique.

As an example consider again the transport of the electrons from the lead 
``a'' to the lead ``b'' due to the single-electron tunneling through the 
quantum dot. Instead of the annihilation and creation operators $a_\sigma 
(\pmb{k})$, $b_\sigma (\pmb{k})$ and $a_\sigma ^ + (\pmb{k})$, $b_\sigma 
^ + (\pmb{k})$ for the electrons in the leads in the Schr\"{o}dinger 
representation the S-matrix (6) is expressed in terms of the corresponding 
operators $a_\sigma (\pmb{k},t)$, $b_\sigma (\pmb{k},t)$ and $a_\sigma ^ + 
(\pmb{k},t)$, $b_\sigma ^ + (\pmb{k},t)$ in the interaction representation 
defined according to the formula (\ref{eq2}) with the expression (2) of the operator 
$H_{0}$ . It is easy to verify that
\begin{equation}
\begin{array}{l}
a_\sigma ^ + (\pmb{k},t) = e^{iE_a (\pmb{k})t}\,a_\sigma ^ + (\pmb{k})\,,\qquad
a_\sigma (\pmb{k},t) = e^{ - iE_a (\pmb{k})t}a_\sigma (\pmb{k})\,,\\
b_{\sigma '}^ + (\pmb{{k}'},{t}') = e^{iE_b (\pmb{{k}'}){t}'}\,b_{\sigma' 
}^ + (\pmb{{k}'})\,,\quad
b_{\sigma '} (\pmb{{k}'},{t}') = e^{ - iE_b (\pmb{{k}'}){t}'}b_{\sigma '} 
(\pmb{{k}'})\,,
\end{array}
\end{equation}
From the expression (6) of the $S$-matrix
\[
S = T\bigg\{ {\exp \bigg( { - i\int\limits_{ - \infty }^\infty {H_{int} 
(t)dt} } \bigg)} \bigg\}
\]
with $H_{int}(t)$ being determined by the formulae (4), (\ref{eq2}) and (43)
\begin{equation}
\begin{array}{l}
 H_{int} (t) = \sum\limits_{\pmb k} {\sum\limits_\sigma {\big\{ {\,V_a 
(\pmb{k})\,e^{iE_a (\pmb{k})t}a_\sigma ^ + (\pmb{k})\,c_\sigma (t) + V_a 
(\pmb{k})^ * \,e^{ - iE_a (\pmb{k})t}c_\sigma ^ + (t)\,a_\sigma 
(\pmb{k})} } } \; \\ 
 \quad \quad \quad \quad \quad \qquad + V_b (\pmb{k})\,e^{iE_b 
(\pmb{k})t}b_\sigma ^ + (\pmb{k})\,c_\sigma (t) + V_b (\pmb{k})^ * \,e^{ 
- iE_b (\pmb{k})t}c_\sigma ^ + (t)\,b_\sigma  {(\pmb{k})} \big\} 
\; \\ 
 \end{array}
\end{equation}
it follows that
\begin{equation}
\begin{array}{l}
 \big\langle {b_{\sigma '} (\pmb{{k}'})\left| R \right|a_\sigma 
(\pmb{k})} \big\rangle = \\ 
\quad \quad - \;iV_a (\pmb{k})^ * V_b (\pmb{{k}'})\int\limits_{ - 
\infty }^\infty {d{t}'} \int\limits_{ - \infty }^\infty {dt\,e^{i[E_b 
(\pmb{{k}'})\,{t}' - E_a (\pmb{k})\,t]}\big\langle {0\,\left| 
{T\{c_{\sigma '} ({t}')\,c_\sigma ^ + (t)S\}\left| 0 \right.} \right.} 
\big\rangle } \,, 
 \end{array}
\end{equation}
\begin{equation}
\begin{array}{l}
 \big\langle {c_{{\sigma }'_2 } b_{{\sigma }'_1 } (\pmb{{k}'})\left| R 
\right|a_{\sigma _1 } (\pmb{k})\,c_{\sigma _2 } } \big\rangle = \\ 
 \quad - \;i\;V_a (\pmb{k})^ * \,V_b (\pmb{{k}'})\int\limits_{ - \infty 
}^\infty {d{t}'} \int\limits_{ - \infty }^\infty {dt\,e^{i[E_b 
(\pmb{{k}'})\,{t}' - E_a (\pmb{k})\,t]}\big\langle {c_{{\sigma }'_2 } 
\big| {T\{c_{{\sigma }'_1 } ({t}')\,c_{\sigma _1 }^ + (t)S\}\big| 
{c_{\sigma _2 } } }} \big\rangle }\,,  
 \end{array}
\end{equation}
\begin{equation}
\begin{array}{l}
 \big\langle {c_ \downarrow c_ \uparrow b_{\sigma '} (\pmb{{k}'})\left| R 
\right|a_\sigma (\pmb{k})\,c_ \uparrow c_ \downarrow } \big\rangle = \\ 
 \quad - \;i\;V_a (\pmb{k})^ * \,V_b (\pmb{{k}'})\int\limits_{ - \infty 
}^\infty {d{t}'} \int\limits_{ - \infty }^\infty {dt\,e^{i[E_b 
(\pmb{{k}'})\,{t}' - E_a (\pmb{k})\,t]}\big\langle {c_ \downarrow c_ 
\uparrow \left| {T\{c_{\sigma '} ({t}')\,c_\sigma ^ + (t)S\}\left| {c_ 
\uparrow c_ \downarrow } \right.} \right.} \big\rangle } . \\ 
 \end{array}
\end{equation}

\n The calculation of these matrix elements (45)\,-\,(47) requires the 
determination of following quantities
\begin{equation}
\label{eq8}
G_{{\sigma }'\sigma }^{(0)cc} ({t}' - t) = - i\big\langle {0\,\left| 
{T\left\{ {\,c_{\sigma '} ({t}')\,c_\sigma ^ + (t)S\,} \right\}\,} 
\right|\,0} \big\rangle \;,
\end{equation}
\begin{equation}
G_{{\sigma }'_2 {\sigma }'_1 \sigma _1 \sigma _2 }^{(\ref{eq1})cc} ({t}' - t) = - 
i\big\langle {c_{{\sigma }'_2 } \big| {T\big\{ {\,c_{{\sigma }'_1 } 
({t}')\,c_{\sigma _1 }^ + (t)\,S\,} \big\}\,\big| {c_{\sigma _2 } } 
} } \big\rangle \,,
\end{equation}
\begin{equation}
\label{eq9}
G_{{\sigma }'\sigma }^{(2)cc} ({t}' - t) = - i\big\langle {c_ \downarrow c_ 
\uparrow \big| {T\,\big\{ {\,c_{\sigma '} ({t}')\,c_\sigma ^ + (t)\,S\,} 
\big\}\,} \big|\,c_ \uparrow c_ \downarrow } \big\rangle 
\end{equation}
which might be called the (generalized) Green functions of the electrons in 
the dot in the presence of the tunneling transitions to and from the leads. 
The expressions of these (generalized) Green functions may be derived by 
means of the Green functions technique.

In order to apply the Green function technique we must work in the 
Heisenberg representation. The transformation from the interaction 
representation to the Heisenberg one is performed with the use of the 
operator $S(t,t_{0})$ satisfying the Schr\"{o}dinger equation with the 
interaction Hamiltonian $H_{int}(t)$, 
\begin{equation}
\label{eq10}
i\frac{d{\kern 1pt} S(t,t_0 )}{dt} = H_{int} (t)\,S(t,t_0 ),
\end{equation}
and the initial condition
\begin{equation}
\label{eq11}
S(t_0 ,t_0 ) = 1.
\end{equation}
The solution of the equation (\ref{eq10}) with the initial condition (\ref{eq11}) can be 
represented in the form similar to the formula (6) for the S-matrix:
\begin{equation}
\label{eq12}
S(t^\prime,t_0 ) = T\bigg\{ {\exp \bigg( { - i\int\limits_{t_0 }^{t^\prime} {H_{int} 
(t)dt} } \bigg)} \bigg\}\, . 
\end{equation}
It is obvious that 
\begin{equation}
\label{eq13}
S = S(\infty , - \infty )\,.
\end{equation}

From the expression (\ref{eq12}) it follows the group property of the operators 
$S(t,t_{0})$:
\begin{equation}
\label{eq14}
S(t,t_0 ) = S(t,{t}')\,S({t}',t_0 )\,.
\end{equation}
From this property and the initial condition (\ref{eq11}) we obtain the relation 
\begin{equation}
\label{eq15}
S(t_0 ,{t}) = S({t},t_0 )^{ - 1}.
\end{equation}
Moreover, the operator (\ref{eq12}) is unitary
\begin{equation}
\label{eq16}
S({t},t_0 )^ + = S({t},t_0 )^{ - 1} = S(t_0 ,{t})\,.
\end{equation}

Now we derive the explicit expression of the chronological product $T\left\{ 
{\,c_{\sigma '} ({t}')\,c_\sigma ^ + (t)\,S\;} \right\}$ in the matrix 
elements (\ref{eq8})\,-\,(\ref{eq9}). In the case $t' > t$ we write
\[
S = S(\infty , - \infty ) = S(\infty ,{t}')\,S({t}',t)\,S(t, - \infty )
\]
and have \sat
$$\begin{array}{l}
 T\big\{ {c_{\sigma '} ({t}')\,c_\sigma ^ + (t)\,S} \big\} = S(\infty 
,{t}')\,c_{\sigma '} ({t}')\,S({t}',t)\,c_\sigma ^ + (t)\,S(t, - \infty ) \\ 
 \quad \quad \quad \quad \quad \quad\quad\quad \;\; = S(\infty , - \infty )\,S( - 
\infty ,{t}')\,c_{\sigma '} ({t}')\,S({t}', - \infty )\,S( - \infty 
,t)c_\sigma ^ + (t)\,S(t, - \infty )\;. \\ 
\end{array}$$
Introducing new operators
\begin{equation}
\label{eq17}
\begin{array}{l}
 C_{\sigma '} ({t}') = S( - \infty ,{t}')\,c_{\sigma '} ({t}')\,S({t}', - 
\infty ), \\ 
 C_\sigma ^ + (t) = S( - \infty ,t)\,c_\sigma ^ + (t)\,S(t, - \infty )\; \\ 
 \end{array}
\end{equation}

\noindent
we rewrite
\sat
\begin{equation}
\label{eq18}
T\big\{ {c_{\sigma '} ({t}')\,c_\sigma ^ + (t)\,S} \big\} = 
S\,C_{\sigma '} ({t}')\,C_\sigma ^ + (t) ,\quad {t}' > t\,.
\end{equation}
Similarly, in the case $t> t'$ we have
\begin{equation}
\label{eq19}
T\big\{ {c_{\sigma '} ({t}')\,c_\sigma ^ + (t)\,S} \big\} = - 
S\,C_\sigma ^ + (t)\;C_{\sigma '} ({t}') ,\quad t > {t}'\,.
\end{equation}
Combining both expressions (\ref{eq18}) and (\ref{eq19}) finally we obtain the formula
\begin{equation}
\label{eq20}
T\big\{ { c_{\sigma '} ({t}')\,c_\sigma ^ + (t)\,S} \big\} = 
ST\big\{ C_{\sigma '} ({t}')\,C_\sigma ^ + (t)\big\}\,.
\end{equation}
Now the (generalized) Green functions (\ref{eq8})\,-\,(\ref{eq9}) are expressed in terms of 
the new operators (\ref{eq17}):
\begin{equation}
\label{eq21}
G_{{\sigma '}\sigma }^{(0)cc} ({t}' - t) = - i\big\langle {0\,\big| 
{ST\big\{ {C_{\sigma '} ({t}')\,C_\sigma ^ + (t)\,} \big\}} 
\big| 0} \big\rangle \,,
\end{equation}

\begin{equation}
G_{{\sigma }'_2 {\sigma }'_1 \sigma _1 \sigma _2 }^{(\ref{eq1})cc} ({t}' - t) = - 
i\big\langle {c_{{\sigma }'_2 } \big| {ST\big\{ {\,C_{{\sigma }'_1 } 
({t}')\,C_{\sigma _1 }^ + (t)} \big\}\big| {c_{\sigma _2 } } } 
} \big\rangle \,,
\end{equation}
\begin{equation}
G_{{\sigma }'\sigma }^{(2)cc} ({t}' - t) = - i\big\langle {c_ \downarrow c_ 
\uparrow \big| {ST\big\{ {C_{\sigma '} ({t}')\,C_\sigma ^ + (t)} 
\big\}} \big| c_ \uparrow c_ \downarrow } \big\rangle \,.
\end{equation}

In order to derive the differential equations for the (generalized) Green 
functions (62)\,-\,(64) we must use the Heisenberg equations of motion for 
the new operators (\ref{eq17}). Denoting
\begin{equation}
\label{eq22}
H(t) = S( - \infty ,t)\,[H_0 + H_{int} (t)]\,S(t, - \infty )\,
\end{equation}
the total Hamiltonian in the new representation we obtain the Heisenberg 
equations of motion with the total Hamiltonian H($t)$:
\begin{equation}
\label{eq23}
i\frac{dC_\sigma (t)}{dt} = [C_\sigma (t),H\,(t)]\,,\quad 
i\frac{dC_\sigma ^ + (t)}{dt} = [C_\sigma ^ + (t),H\,(t)]\,.
\end{equation}
This means that $C_\sigma (t)$ and $C_\sigma ^ + (t)$ are the annihilation 
and creation operators in the Heisenberg representation.

Together with the operators $C_\sigma (t)$ and $C_\sigma ^ + (t)$ determined 
by the formula (\ref{eq17}) introduce also the annihilation and creation operators 
for the electrons in the leads in the Heisenberg representation

\begin{equation}
\label{eq24}
\begin{array}{l}
 A_\sigma (\pmb{k},t) = S( - \infty ,t)\,a_\sigma (\pmb{k},t)\,S(t, - 
\infty )\,, \\ 
 A_\sigma ^ + (\pmb{k},t) = S( - \infty ,t)\,a_\sigma ^ + 
(\pmb{k},t)\,S(t, - \infty )\,, \\ 
 B_\sigma (\pmb{k},t) = S( - \infty ,t)\,b_\sigma (\pmb{k},t)\,S(t, - 
\infty )\, , \\ 
 B_\sigma ^ + (\pmb{k},t) = S( - \infty ,t)\,b_\sigma ^ + 
(\pmb{k},t)\,S(t, - \infty )\,. \\ 
 \end{array}
\end{equation}
They satisfy also the Heisenberg equations of motion with the total 
Hamiltonian H($t)$:
\sat

\begin{equation}
\label{eq25}
\begin{array}{l}
 i\dfrac{dA_\sigma (\pmb{k},t)}{dt} = [A_\sigma (\pmb{k},t),H\,(t)]\, 
,\quad \quad i\dfrac{dA_\sigma ^ + (\pmb{k},\pmb{ }t)}{dt} = [A_\sigma ^ + 
(\pmb{k},t),H\,(t)]\, , \\ 
 i\dfrac{dB_\sigma (\pmb{k},t)}{dt} = [B_\sigma (\pmb{k},t),H\,(t)]\, 
,\quad \quad i\dfrac{dB_\sigma ^ + (\pmb{k},\pmb{ }t)}{dt} = [B_\sigma ^ + 
(\pmb{k},t),H\,(t)]\,. \\ 
 \end{array}
\end{equation}

The total Hamiltonian in the Heisenberg representation is expressed directly 
in terms of the new operators (\ref{eq17}) and (\ref{eq24}):
\sat
\begin{equation}
\label{eq26}
\begin{array}{l}
 H\,(t)\; = \; E\sum\limits_\sigma {C_\sigma ^ + (t)\,C_\sigma (t) + UN_ 
\uparrow (t)N_ \downarrow (t) + } \\ 
 \qquad \quad \quad + \sum\limits_{\pmb k} {\sum\limits_\sigma {\left\{ {E_a 
(\pmb{k})A_\sigma ^ + (\pmb{k},t)\,A_\sigma (\pmb{k},t) + E_b 
(\pmb{k})B_\sigma ^ + (\pmb{k},t)B_\sigma (\pmb{k},t\left. ) \right\} } 
\right.} } + \\ 
 \qquad \quad \quad + \sum\limits_{\pmb k} {\sum\limits_\sigma {\left\{ {\;V} 
\right._a (\pmb{k})A_\sigma ^ + (\pmb{k},t)\,C_\sigma (t) + V_a 
(\pmb{k})^ * C_\sigma ^ + (t)\,A_\sigma (\pmb{k},t) + } } \\ 
 \quad \quad \quad \quad \quad \quad + V_b (\pmb{k})B_\sigma ^ + 
(\pmb{k},t)\,C_\sigma (t) + V_b (\pmb{k})^ * C_\sigma ^ + (t)\,B_\sigma 
(\pmb{k},t\left. ) \right\} \, , \\ 
 \end{array}
\end{equation}

\[
N_\sigma (t) = C_\sigma ^ + (t)\;C_\sigma (t)\,.
\]

\n 
The operators (\ref{eq17}) and (\ref{eq24}) satisfy the canonical equal-time anticommutation 
relations

\begin{equation}
\label{eq27}
\begin{array}{l}
 \{C_\sigma (t)\;,\;C_{\sigma '}^ + (t)\} = \delta _{\sigma {\sigma }'} \,, 
\\ 
 \{A_\sigma  (\pmb{k},t)\;,\;A_{\sigma '}^ + (\pmb{{k}'},t)\} = \delta 
_{\sigma {\sigma }'} \,\delta _{\pmb{k{k}'}} \,, \\ 
 \{B_\sigma (\pmb{k},t)\;,\;B_{\sigma '}^ + (\pmb{{k}'},t)\} = \delta 
_{\sigma {\sigma }'} \,\delta _{\pmb{k{k}'}} \, , \\ 
 \end{array}
\end{equation}
other equal-time anticommutators between these operators being equal to 
zero. 

From the Heisenberg equations of motion (\ref{eq23}) and (\ref{eq25}), the expression (\ref{eq26}) 
of the total Hamiltonian, the equal-time canonical anticommutation relations 
(\ref{eq27}) and other ones with the vanishing r. h. s. we derive following 
differential equations for the operators (\ref{eq17}) and (\ref{eq24}):
\begin{equation} 
i\frac{dC_\sigma (t)}{dt} = EC_\sigma (t) + UN_{ - \sigma } (t)\;C_\sigma 
(t) + \sum\limits_{\pmb k} {\left\{ {\,V_a (\pmb{k})^ * A_\sigma 
(\pmb{k},t) + V_b (\pmb{k})^ * B_\sigma (\pmb{k},t)} \right\},} 
\end{equation}
\begin{equation} 
 i\frac{dA_\sigma (\pmb{k},t)}{dt} = E_a (\pmb{k})A_\sigma (\pmb{k},t) + 
V_a (\pmb{k})\,C_\sigma (t)\,, 
\end{equation}
\begin{equation}
 i\frac{dB_\sigma (\pmb{k},t)}{dt} = E_b (\pmb{k})B_\sigma (\pmb{k},t) + 
V_b (\pmb{k})\,C_\sigma (t)\,. 
\end{equation}
Using these differential equations and the canonical equal-time 
anticommutation relations we can derive the systems of equations for the 
(generalized) Green functions.

\vss

\ce{\bf IV. CALCULATION OF THE (GENERALIZED) GREEN FUNCTIONS}

\vs
For the convenience in the determination of the real and imaginary parts of 
the Fourier transforms of the (generalized) Green functions (\ref{eq21})--(64) we 
divide each of them into two functions according to the definition of the 
chronological product
\[
T\left\{ {\,C_{\sigma '}^ ({t}')\,C_\sigma ^ + (t)\,} \right\} = \theta 
({t}' - t)\,C_{\sigma'} ({t}')C_\sigma ^ + (t) - \theta (t - 
{t}')\,C_\sigma ^ + (t)\,C_{\sigma '} ({t}')\,.
\]
We have
\begin{equation}
G_{{\sigma }'\sigma }^{(0)cc} ({t}' - t) = G_{{\sigma }'\sigma }^{(0)\,cc( + 
)} ({t}' - t) + G_{{\sigma }'\sigma }^{(0)\,cc( - )} ({t}' - t)\,,
\end{equation}
\begin{equation}
G_{{\sigma }'_2 {\sigma }'_1 \sigma _1 \sigma _2 }^{(\ref{eq1})cc} ({t}' - t) = 
G_{{\sigma }'_2 {\sigma }'_1 \sigma _1 \sigma _2 }^{(\ref{eq1})cc( + )} ({t}' - t) + 
G_{{\sigma }'_2 {\sigma }'_1 \sigma _1 \sigma _2 }^{(\ref{eq1})\;cc( - )} ({t}' - 
t)\,,
\end{equation}
\begin{equation}
G_{{\sigma }'\sigma }^{(2)cc} ({t}' - t) = G_{{\sigma }'\sigma }^{(2)cc( + 
)} ({t}' - t) + G_{{\sigma }'\sigma }^{(2)cc( - )} ({t}' - t)
\end{equation}
with
\sat
\begin{eqnarray}
&& G_{{\sigma }'\sigma }^{(0)cc( + )} ({t}' - t) = - i\theta ({t}' - 
t)\big\langle {0\,\big| {S\,C_{\sigma '} ({t}')\,C_\sigma ^ + (t)\,} 
\big|\,0} \big\rangle \,, \\ 
&& G_{{\sigma }'\sigma }^{(0)cc( - )} ({t}' - t) = i\theta (t - 
{t}')\big\langle {0\big| {S\,C_\sigma ^ + (t)\,C_{\sigma '} ({t}')\,} 
\big|\,0} \big\rangle \,, \\ 
&& G_{{\sigma }'_2 {\sigma }'_1 \sigma _1 \sigma _2 }^{(\ref{eq1})cc( + )} ({t}' - t) 
= - i\theta ({t}' - t)\big\langle {c_{{\sigma }'_2 } \big| {SC_{{\sigma 
}'_1 } ({t}')C_{\sigma _1 }^ + (t)\big| {c_{\sigma _2 } } } 
} \big\rangle \,, \\ 
&& G_{{\sigma }'_2 {\sigma '}_1 \sigma _1 \sigma _2 }^{(\ref{eq1})cc( - )} ({t}' - t) 
= i\theta (t - {t}')\big\langle {c_{{\sigma '}_2 } \big| {SC_{\sigma _1 }^ 
+ (t)C_{{\sigma }'_1 } ({t}')\big| {c_{\sigma _2 } } } } 
\big\rangle \,, \\ 
 && G_{{\sigma }'\sigma }^{(2)cc( + )} ({t}' - t) = - i\theta ({t}' - 
t)\big\langle {c_ \uparrow c_ \downarrow \big| {SC_{\sigma '} 
({t}')\,C_\sigma ^ + (t)} \big|c_ \downarrow c_ \uparrow } 
\big\rangle \,, \\ 
&& G_{{\sigma }'\sigma }^{(2)cc( - )} ({t}' - t) = i\theta (t - 
{t}')\big\langle {c_ \uparrow c_ \downarrow \big| {SC_\sigma ^ + 
(t)C_{\sigma '} ({t}')} \big| c_ \downarrow c_ \uparrow } \big\rangle 
\,.
\end{eqnarray}

Using the equation (71) for 
$C_{\sigma '} ({t}')$ we can derive the differential equations for the 
functions $G_{{\sigma }'\sigma }^{(0)cc(\pm )} ({t}' - t)\;,\;G_{{\sigma }'_2 {\sigma 
}'_1 \sigma _1 \sigma _2 }^{(1)cc(\pm )} ({t}' - t)\;$
and $G_{{\sigma }'\sigma }^{(2)cc(\pm )} ({t}' - t)$ 
which contain many other (generalized) Green functions. The 
derived differential equations do not form closed finite systems of 
equations, but belong to some infinite systems. In order to find their approximate 
solutions we might truncate them at a corresponding order and obtain the 
closed systems of approximate differential equations which can be solved 
exactly. Let us now truncate the infinite systems of equations by neglecting 
the 4-point (generalized) Green functions multiplied with small functions 
$V_a (\pmb{k}),\;V_b (\pmb{k}),\;V_a (\pmb{k})^ * ,\;V_b (\pmb{k})^ * $ 
and all $n$-point (generalized) Green functions with $n > 4$. Then we obtain 
closed systems of approximate equations containing following 
time-independent matrix elements (constants):
\begin{align}
 &\big\langle {0 \left| {S\,C_{\sigma '} (t)C_\sigma ^ + (t)\,} \right| 0} 
\big\rangle  = g_{{\sigma }'\sigma }^{(0)cc( + )} ,\nn \\ 
&\big\langle {0 \left| {S\,C_\sigma ^ + (t) C_{\sigma '} (t)\,} 
\right| 0} \big\rangle  = g_{{\sigma }'\sigma }^{(0)cc( - )} ,\nn \\ 
& \big\langle {c_{{\sigma }'_2 } \big| {S\,C_{{\sigma }'_1 } (t)C_{\sigma 
_1 }^ + (t)} \big| c_{\sigma _2 } } \big\rangle  = g_{{\sigma }'_2 
{\sigma }'_1 \sigma _1 \sigma _2 }^{(\ref{eq1})cc( + )} , \nn\\ 
&\big\langle {c_{{\sigma }'_2 } \big| {S\,C_{\sigma _1 }^ + 
(t)C_{{\sigma }'_1 } (t)\,} \big| c_{\sigma _2 } } \big\rangle  = 
g_{{\sigma }'_2 {\sigma }'_1 \sigma _1 \sigma _2 }^{(\ref{eq1})cc( - )} \,,\nn \\ 
& \big\langle {c_ \downarrow c_ \uparrow \left| {S\,C_{\sigma '} 
(t)C_\sigma ^ + (t)} \right| c_ \uparrow c_ \downarrow } \big\rangle  
= g_{{\sigma }'\sigma }^{(2)cc( + )} ,\nn \\ 
& \big\langle {c_ \downarrow c_ \uparrow \left| {S\,C_\sigma ^ + 
(t)C_{\sigma '} (t)} \right| c_ \uparrow c_ \downarrow } \big\rangle  
= g_{{\sigma }'\sigma }^{(2)cc( - )}\, ,\nn \\ 
& \big\langle {0 \left| {S\,A_{\sigma '} (\pmb{k},{t})\,C_\sigma ^ + 
(t)} \right| 0} \big\rangle = g_{{\sigma }'\sigma }^{(0)ac( + )} 
(\pmb{k})\, , \nn\\ 
& \big\langle {0 \left| {S\,C_\sigma ^ + (t)A_{\sigma '} (\pmb{k},{t})} 
\right| 0} \big\rangle = g_{{\sigma }'\sigma }^{(0)ac( - )} (\pmb{k})\,, \nn\\ 
& \big\langle {c_{{\sigma }'_2 } \big| {S\,A_{{\sigma }'_1 } 
(\pmb{k},t)C_{\sigma _1 }^ + (t) } \big| c_{\sigma _2 } } \big\rangle 
 = g_{{\sigma }'_2 {\sigma }'_1 \sigma _1 \sigma _2 }^{(\ref{eq1})ac( + )} 
(\pmb{k})\,,\\ 
& \big\langle {c_{{\sigma }'_2 } \big| {S\,C_{\sigma _1 }^ + 
(t)A_{{\sigma }'_1 } (\pmb{k},t)} \big| c_{\sigma _2 } } \big\rangle 
= g_{{\sigma }'_2 {\sigma }'_1 \sigma _1 \sigma _2 }^{(\ref{eq1})ac( - )} 
(\pmb{k})\,, \nn\\ 
& \big\langle {c_ \downarrow c_ \uparrow \left| {S\,A_{\sigma '} 
(\pmb{k},t)C_\sigma ^ + (t)} \right| c_ \uparrow c_ \downarrow } 
\big\rangle = g_{{\sigma }'\sigma }^{(2)ac( + )} (\pmb{k}),\nn \\ 
& \big\langle {c_ \downarrow c_ \uparrow \left| {S\,C_\sigma ^ + 
(t)A_{\sigma '} (\pmb{k},t)} \right| c_ \uparrow c_ \downarrow } 
\big\rangle  = g_{{\sigma }'\sigma }^{(2)ac( - )} (\pmb{k})\,,\nn \\ 
& \big\langle {0 \left| {S\,B_{\sigma '} (\pmb{k},{t})\,C_\sigma ^ + 
(t)} \right| 0} \big\rangle = g_{{\sigma }'\sigma }^{(0)bc( + )} 
(\pmb{k})\,, \nn  \\ 
& \big\langle {0 \left| {S\,C_\sigma ^ + (t)B_{\sigma '} (\pmb{k},{t})} 
\right| 0} \big\rangle = g_{{\sigma }'\sigma }^{(0)bc( - )} (\pmb{k})\;, \nn\\
 & \big\langle {c_{{\sigma }'_2 } \big| {S\,B_{{\sigma }'_1 } 
(\pmb{k},t)C_{\sigma _1 }^ + (t)} \big| c_{{\sigma }'_2 } } 
\big\rangle = g_{{\sigma }'_2 {\sigma }'_1 \sigma _1 \sigma _2 
}^{(\ref{eq1})bc( + )} (\pmb{k}),\nn \\ 
& \big\langle {c_{{\sigma }'_2 } \big| {S\,C_{\sigma _1 }^ + 
(t)B_{{\sigma }'_1 } (\pmb{k},t)} \big| c_{{\sigma }'_2 } } 
\big\rangle  = g_{{\sigma }'_2 {\sigma }'_1 \sigma _1 \sigma _2 
}^{(\ref{eq1})bc( - )} (\pmb{k})\,,\nn \\ 
& \big\langle {c_ \downarrow c_ \uparrow \left| {S\,B_{\sigma '} 
(\pmb{k},t)C_\sigma ^ + (t)} \right| c_ \uparrow c_ \downarrow } 
\big\rangle  = g_{{\sigma }'\sigma }^{(2)bc( + )} (\pmb{k})\,,\nn \\ 
& \big\langle {c_ \downarrow c_ \uparrow  \left| {S\,C_\sigma ^ + 
(t)B_{\sigma '} (\pmb{k},t)} \right| c_ \uparrow c_ \downarrow } 
\big\rangle  = g_{{\sigma }'\sigma }^{(2)bc( - )} (\pmb{k})\,,\nn
 \\ 
&\big\langle 0 \left| S\,N_{-{\sigma}'} (t)\,C_{\sigma '} 
(t)C_{\sigma}^{+} (t) \right| 0 \big\rangle = f_{{\sigma }'\sigma 
}^{(0)cc( + )}\,,\nn \\ 
& \big\langle {0 \left| {S\,N_{ - {\sigma }'} (t)\,A_{\sigma '} 
(\pmb{k},{t})\,C_\sigma ^ + (t)} \right| 0} \big\rangle = f_{{\sigma 
}'\sigma }^{(0)ac( + )} (\pmb{k})\,,\nn
\\ 
& \big\langle {0 \big| {S\,N_{ - {\sigma }'} (t)\,B_{\sigma '} 
(\pmb{k},{t})\,C_\sigma ^ + (t)} \big| 0} \big\rangle  = 
f_{{\sigma }'\sigma }^{(0)bc( + )} (\pmb{k})\,, \nn \\
& \big\langle {0 \left| {S\,C_\sigma ^ + (t)\,N_{ - {\sigma }'} 
(t)C_{\sigma '} (t)} \right| 0} \big\rangle = f_{{\sigma 
}'\sigma }^{(0)cc( - )} \,,\nn
 \\ 
& \big\langle {0 \left| {S\,C_\sigma ^ + (t)N_{ - {\sigma }'} 
(t)\,A_{\sigma '} (\pmb{k},{t})} \right| 0} \big\rangle = f_{{\sigma 
}'\sigma }^{(0)ac( - )} (\pmb{k})\,,\nn\\  
& \big\langle {0 \big| {S\,C_\sigma ^ + (t)\,N_{ - {\sigma }'} 
(t)\,B_{\sigma '}^ (\pmb{k},{t})} \big| 0} \big\rangle  = 
f_{{\sigma }'\sigma }^{(0)bc( - )} (\pmb{k})\,,\nn\\
& \big\langle {c_{{\sigma }'_2 } \big| {S\,N_{ - {\sigma }'_1 } 
(t)\,C_{{\sigma }'_1 } (t)C_{\sigma _1 }^ + (t)} \big| c_{\sigma _2 } } 
\big\rangle  = f_{{\sigma }'_2 {\sigma }'_1 \sigma _1 \sigma _2 
}^{(\ref{eq1})cc( + )}\, ,\nn\\
& \big\langle {c_{{\sigma }'_2 } \big| {S\,N_{ - {\sigma }'_1 } 
(t)\,A_{{\sigma }'_1 } (\pmb{k},{t})\,C_{\sigma _1 }^ + (t)} 
\big| c_{\sigma _2 } } \big\rangle = f_{{\sigma }'_2 {\sigma }'_1 
\sigma _1 \sigma _2 }^{(\ref{eq1})ac( + )} (\pmb{k})\,,\nn \\ 
& \big\langle {c_{{\sigma }'_2 }  \big| {S\,N_{ - {\sigma }'_1 } 
(t)\,B_{{\sigma }'_1 } (\pmb{k},{t})\,C_{\sigma _1 }^ + (t)} 
\big| c_{\sigma _2 } } \big\rangle  = f_{{\sigma }'_2 {\sigma }'_1 
\sigma _1 \sigma _2 }^{(\ref{eq1})bc( + )} (\pmb{k})\,,\nn 
\end{align}

\newpage
$\quad$
\vskip-1.2cm
\begin{align}
&\big\langle {c_{{\sigma }'_2 } \big| {S\,C_{\sigma _1 }^ + (t)\,N_{ - 
{\sigma }'_1 } (t)C_{{\sigma }'_1 } (t)} \big| c_{\sigma _2 
} } \big\rangle = f_{{\sigma }'_2 {\sigma }'_1 \sigma _1 \sigma _2 
}^{(\ref{eq1})cc( - )} \,,\nn \\ 
& \big\langle {c_{{\sigma }'_2 } \big| {S\,C_{\sigma _1 }^ + (t)N_{ - 
{\sigma }'_1 } (t)\,A_{{\sigma }'_1 } (\pmb{k},{t})} \big| c_{\sigma 
_2 } } \big\rangle = f_{{\sigma }'_2 {\sigma }'_1 \sigma _1 \sigma _2 
}^{(\ref{eq1})ac( - )} (\pmb{k})\,,\nn \\ 
& \big\langle {c_{{\sigma }'_2 } \big| {S\,C_{\sigma _1 }^ + (t)\,N_{ - 
{\sigma }'_1 } (t)\,B_{{\sigma }'_1 } (\pmb{k},{t})} 
\big| c_{\sigma _2 } } \big\rangle  = f_{{\sigma }'_2 {\sigma }'_1 
\sigma _1 \sigma _2 }^{(\ref{eq1})bc( - )} (\pmb{k})\,,\nn \\ 
& \left\langle {c_ \downarrow c_ \uparrow \,\left| {S\,N_{ - {\sigma }'} 
(t)\,C_{\sigma '} (t)C_\sigma ^ + (t)\,} \right|\,c_ \uparrow c_ \downarrow 
} \right\rangle  = f_{{\sigma }'\sigma }^{(2)cc( + )}\, ,\nn \\ 
&\left\langle {c_ \downarrow c_ \uparrow  \left| {S\,N_{ - {\sigma }'}
(t)\,A_{\sigma '} (\pmb{k},{t})\,C_\sigma ^ + (t)} \right|c_ \uparrow 
c_ \downarrow } \right\rangle = f_{{\sigma }'\sigma }^{(2)ac( + )} 
(\pmb{k})\,,\nn \\ 
&\left\langle {c_ \downarrow c_ \uparrow \left| {S\,N_{ - {\sigma }'}
(t)\,B_{\sigma '} (\pmb{k},{t})\,C_\sigma ^ + (t)}  \right| c_ \uparrow 
c_ \downarrow } \right\rangle  = f_{{\sigma }'\sigma }^{(2)bc( + )} 
(\pmb{k})\,,\nn \\ 
& \left\langle {c_ \downarrow c_ \uparrow \left| {S\,C_\sigma ^ + (t)\,N_{ 
- {\sigma }'} (t)C_{\sigma '} (t) } \right| c_ \uparrow c_ 
\downarrow } \right\rangle = f_{{\sigma }'\sigma }^{(2)cc( - )} \;,\nn \\ 
& \left\langle {c_ \downarrow c_ \uparrow \left| {S\,C_\sigma ^ + (t)N_{ - 
{\sigma }'} (t)\,A_{\sigma '} (\pmb{k},{t})} \right| c_ \uparrow c_ 
\downarrow } \right\rangle = f_{{\sigma }'\sigma }^{(2)ac( - )} 
(\pmb{k})\,, \nn\\ 
& \big\langle {c_ \downarrow c_ \uparrow  \big| {S\,C_\sigma ^ + (t)\,N_{ 
- {\sigma }'} (t)\,B_{\sigma '} (\pmb{k},{t})} \big| c_ \uparrow c_ 
\downarrow } \big\rangle  = f_{{\sigma }'\sigma }^{(2)bc( - )} 
(\pmb{k})\,. \nn 
\end{align}

Performing the Fourier transformation of the differential equations we 
derive a system of algebraic equations which can be solved easily and obtain 
following results:
\begin{align}
& \tilde {G}_{\sigma'\sigma }^{(0)cc(\pm )} (\omega ) = \{\omega - E - 
\Sigma (\,\omega \,\pm io)\}^{ - 1}\, \cdot \\ 
& \quad \quad\quad \left\{ {g_{\sigma' \sigma }^{(0)cc(\pm )} + 
\sum\limits_{{\pmb k}} {\left[ {\frac{g_{{\sigma }' \sigma }^{(0)ac(\pm )} 
({\pmb k})}{\omega  - E_a ({\pmb k})\pm io} + \frac{g_{{\sigma }' 
\sigma }^{(0)bc(\pm )} ({\pmb k})}{\omega \mbox{ - }E_b ({\pmb k})\pm io}} 
\right] + U\tilde {F}_{{\sigma }' \sigma }^{(0)cc(\pm )} (\omega )} } 
\right\}\,,\nn 
\end{align}
\begin{align}
& \tilde {G}_{{\sigma }'_2 {\sigma }'_1 \sigma _1 \sigma _2 }^{(\ref{eq1})cc(\pm )} 
(\omega ) = \{\omega - E - \Sigma (\,\omega \,\pm io)\}^{ - 1} \, \cdot \\ 
&\quad \quad \quad \left\{ {g_{{\sigma }'_2 {\sigma }'_1 \sigma _1 
\sigma _2 }^{(\ref{eq1})cc(\pm )} + \sum\limits_{\pmb k} {\left[ {\frac{g_{{\sigma 
}'_2 {\sigma }'_1 \sigma _1 \sigma _2 }^{(\ref{eq1})ac(\pm )} ({\pmb k})}{\omega 
\mbox{ - }E_a ({\pmb k})\pm io} + \frac{g_{{\sigma }'_2 {\sigma }'_1 \sigma 
_1 \sigma _2 }^{(\ref{eq1})bc(\pm )} ({\pmb k})}{\omega \mbox{ - }E_b ({\pmb k})\pm 
io}} \right] + U\tilde {F}_{{\sigma }'_2 {\sigma }'_1 \sigma _1 \sigma _2 
}^{(\ref{eq1})cc(\pm )} (\omega )} } \right\}\,,\nn 
\end{align}
\begin{align}
&\tilde {G}_{\sigma'\sigma }^{(2)cc(\pm )} (\omega ) = \{\omega - E - 
\Sigma (\,\omega \,\pm io)\}^{ - 1} \, \cdot \\ 
& \quad \quad \quad \left\{ {g_{\sigma '\sigma}^{(2)cc(\pm )} + 
\sum\limits_{\pmb k} {\left[ {\frac{g_{{\sigma }' \sigma }^{(2)ac(\pm )} 
({\pmb k})}{\omega \mbox{ - }E_a ({\pmb k})\pm io} + \frac{g_{{\sigma }' 
\sigma }^{(2)bc(\pm )} ({\pmb k})}{\omega \mbox{ - }E_b ({\pmb k})\pm io}} 
\right] + U\tilde {F}_{{\sigma }' \sigma }^{(2)cc(\pm )} (\omega )} } 
\right\}\,, \nn
\end{align}
\begin{align}
&\tilde {F}_{{\sigma }'\sigma }^{(0)cc(\pm )} (\omega ) = \{\omega - E - U - 
\Sigma (\,\omega \pm io)\}^{ - 1} \, \cdot \\ 
&\quad \quad \quad \left\{ {f_{{\sigma }'\sigma 
}^{(0)cc(\pm )} + \sum\limits_{\pmb k} {\left[ {\frac{f_{{\sigma }'\sigma 
}^{(0)ac(\pm )} ({\pmb k})}{\omega \mbox{ - }E_a ({\pmb k})\pm io} + 
\frac{f_{{\sigma }'\sigma }^{(0)bc(\pm )} ({\pmb k})}{\omega \mbox{ - }E_b 
({\pmb k})\pm io}} \right]} } \right\}\,, \nn 
 \end{align}
\begin{align}
& \tilde {F}_{{\sigma }'_2 {\sigma }'_1 \sigma _1 \sigma _2 }^{(\ref{eq1})cc(\pm )} 
(\omega ) = \{\omega - E - U - \Sigma (\,\omega \,\pm io)\}^{ - 1} \, \cdot \\ 
&\quad \quad \quad \left\{ {f_{{\sigma }'_2 {\sigma 
}'_1 \sigma _1 \sigma _2 }^{(\ref{eq1})cc(\pm )} + \sum\limits_{\pmb k} {\left[ 
{\frac{f_{{\sigma }'_2 {\sigma }'_1 \sigma _1 \sigma _2 }^{(\ref{eq1})ac(\pm )} 
({\pmb k})}{\omega \mbox{ - }E_a ({\pmb k})\pm io} + \frac{f_{{\sigma }'_2 
{\sigma }'_1 \sigma _1 \sigma _2 }^{(\ref{eq1})bc(\pm )} ({\pmb k})}{\omega \mbox{ - 
}E_b ({\pmb k})\pm io}} \right]} } \right\}\,, \nn
 \end{align}
\pagebreak

$\quad$
\vskip-1.3cm
\begin{align}
& \tilde {F}_{{\sigma }' \sigma }^{(2)cc(\pm )} (\omega ) = \{\omega - E - U 
- \Sigma (\,\omega \,\pm io)\}^{ - 1} \,\cdot \\ 
& \quad \quad \quad \left\{ {f_{{\sigma }' \sigma 
}^{(2)cc(\pm )} + \sum\limits_{\pmb k} {\left[ {\frac{f_{{\sigma }' \sigma 
}^{(2)ac(\pm )} ({\pmb k})}{\omega \mbox{ - }E_a ({\pmb k})\pm io} + 
\frac{f_{{\sigma }' \sigma }^{(2)bc(\pm )} ({\pmb k})}{\omega \mbox{ - }E_b 
({\pmb k})\pm io}} \right]} } \right\}\,, \nn
 \end{align}
where
\vskip-0.7cm
\begin{align}
\Sigma\,\omega \pm io) = \sum\limits_{\pmb k} {\left[ {\frac{\vert V_a 
({\pmb k})\vert ^2}{\omega \mbox{ - }E_a ({\pmb k})\pm io} + \frac{\vert V_b 
({\pmb k})\vert ^2}{\omega \mbox{ - }E_b ({\pmb k})\pm io}} \right]} \;
\end{align}
is the self-energy part of the electron in the dot due to its tunneling from 
and to the leads. The self-energy part $\Sigma (\omega )$ 
in the denominators of the (generalized) Green functions describes
the broadening of the energy levels of the electron in the dot.

The constants (83) can be calculated in the perturbation theory with respect 
to the tunnelling Hamiltonian $H_{int}$. In the lowest (zero) order we have 
following values:
\begin{align}
&f_{{\sigma }'\sigma }^{(0)cc( \pm )} = f_{{\sigma }'\sigma }^{(0)ac(\pm )} 
({\pmb k}) = f_{{\sigma }'\sigma }^{(0)bc(\pm )} ({\pmb k}) = 0\,,\nn \\ 
& f_{{\sigma }'_2 {\sigma }'_1 \sigma _1 \sigma _2 }^{(\ref{eq1})cc( + )} = \delta _{ 
- \sigma _1 \sigma _2 } \delta _{ - {\sigma }'_1 {\sigma }'_2 } (\delta 
_{{\sigma }'_1 \sigma _1 } - \delta _{{\sigma }'_1 \sigma _2 } )\,,\quad 
f_{{\sigma }'_2 {\sigma }'_1 \sigma _1 \sigma _2 }^{(\ref{eq1})cc( - )} = 0\,,\nn\\ 
& f_{{\sigma }'_2 {\sigma }'_1 \sigma _1 \sigma _2 }^{(\ref{eq1})ac(\pm )} ({\pmb k}) 
= f_{{\sigma }'_2 {\sigma }'_1 \sigma _1 \sigma _2 }^{(\ref{eq1})bc( + )} ({\pmb k}) 
= 0\,, \nn \\ 
& f_{{\sigma }'\sigma }^{(2)cc( + )} = 0\, ,\quad f_{{\sigma 
}'\sigma }^{(2)cc( - )} = \delta _{{\sigma }'\sigma } \, ,\nn \\ 
& f_{{\sigma }'\sigma }^{(2)ac(\pm )} ({\pmb k}) = f_{{\sigma }'\sigma 
}^{(2)bc(\pm )} ({\pmb k}) = 0\,,\\ 
& g_{{\sigma }'\sigma }^{(0)cc( + )} = \delta _{{\sigma }'\sigma } \, 
,\quad g_{{\sigma }'\sigma }^{(0)cc( - )} = 0\, ,\nn \\ 
& g_{{\sigma }'\sigma }^{(0)ac(\pm )} ({\pmb k}) = g_{{\sigma }'\sigma 
}^{(0)bc(\pm )} ({\pmb k}) = 0\,,\nn \\ 
& g_{{\sigma }'_2 {\sigma }'_1 \sigma _1 \sigma _2 }^{(\ref{eq1})cc( + )} = \delta _{ 
- \sigma _1 \sigma _2 } \delta _{ - {\sigma }'_1 {\sigma }'_2 } (\delta 
_{{\sigma }'_1 \sigma _1 } - \delta _{{\sigma }'_1 \sigma _2 } )\,, \nn \\ 
& g_{{\sigma }'_2 {\sigma }'_1 \sigma _1 \sigma _2 }^{(\ref{eq1})cc( - )} = \delta 
_{\sigma _1 {\sigma }'_2 } \delta _{\sigma _2 {\sigma }'_1 } \,,\nn \\ 
& g_{{\sigma }'_2 {\sigma }'_1 \sigma _1 \sigma _2 }^{(\ref{eq1})ac(\pm )} ({\pmb k}) 
= g_{{\sigma }'_2 {\sigma }'_1 \sigma _1 \sigma _2 }^{(\ref{eq1})bc( \pm )} ({\pmb k}) 
= 0\,,\nn \\ 
& g_{{\sigma }'\sigma }^{(2)cc( + )} = 0\,,\quad g_{{\sigma 
}'\sigma }^{(2)cc( - )} = \delta _{{\sigma }'\sigma } \quad ,\nn \\ 
& g_{{\sigma }'\sigma }^{(2)ac(\pm )} ({\pmb k}) = g_{{\sigma }'\sigma 
}^{(2)bc(\pm )} ({\pmb k}) = 0\,.\nn 
\end{align}
Substituting these values of the constants into the r. h. s. of the 
relations (84)\,-\,(89) we obtain
\begin{align}
& \tilde {F}_{{\sigma }'\sigma }^{(0)cc(\pm )} (\omega ) = 0\, , \\ 
& \tilde {F}_{{\sigma }'_2 {\sigma }'_1 \sigma _1 \sigma _2 }^{(\ref{eq1})cc( + )} 
(\omega ) = \frac{\delta _{ - \sigma _1 \sigma _2 } \delta _{ - {\sigma }'_1 
{\sigma }'_2 } (\delta _{{\sigma }'_1 \sigma _1 } - \delta _{{\sigma }'_1 
\sigma _2 } )}{\omega - E - U - \Sigma (\omega + io)}\,,\\ 
&\tilde {F}_{{\sigma }'_2 {\sigma }'_1 \sigma _1 \sigma _2 }^{(\ref{eq1})cc( - )} 
(\omega ) = 0\,, \\
& \tilde {F}_{{\sigma }'\sigma }^{(2)cc( + )} (\omega ) = 0\,, \\ 
& \tilde {F}_{{\sigma }'\sigma }^{(2)cc( - )} (\omega ) = \frac{\delta 
_{{\sigma }'\sigma } }{\omega - E - U - \Sigma (\omega - io)}\,,
\end{align} 
\pagebreak

$\quad$
\vskip-1.2cm
\begin{align}
& \tilde {G}_{{\sigma }'\sigma }^{(0)cc( + )} (\omega ) = \frac{\delta 
_{{\sigma }'\sigma } }{\omega - E - \Sigma (\omega + io)}\,, \\ 
& \tilde {G}_{{\sigma }'\sigma }^{(0)cc( - )} (\omega ) = 0\,, \\ 
& \tilde {G}_{{\sigma }'_2 {\sigma }'_1 \sigma _1 \sigma _2 }^{(\ref{eq1})cc( + )} 
(\omega ) = \frac{\delta _{ - \sigma _1 \sigma _2 } \delta _{ - {\sigma }'_1 
{\sigma }'_2 } (\delta _{{\sigma }'_1 \sigma _1 } - \delta _{{\sigma }'_1 
\sigma _2 } )}{\omega - E - U - \Sigma (\omega + io)}\,, \\ 
& \tilde {G}_{{\sigma }'_2 {\sigma }'_1 \sigma _1 \sigma _2 }^{(\ref{eq1})cc( - )} 
(\omega ) = \frac{\delta _{\sigma _1 {\sigma }'_2 } \delta _{\sigma _2 
{\sigma }'_1 } }{\omega - E - \Sigma (\omega - io)}\,, \\ 
& \tilde {G}_{{\sigma }'\sigma }^{(2)cc( + )} (\omega ) = 0\,, \\ 
& \tilde {G}_{{\sigma }'\sigma }^{(2)cc( - )} (\omega ) = \frac{\delta 
_{{\sigma }'\sigma } }{\omega - E - U - \Sigma (\omega - io)}\, \cdot 
\end{align} 
Summing up the two parts of each (generalized) Green functions $\tilde 
{G}_{{\sigma }'\sigma }^{(0)cc} (\omega )\;,\;\;\tilde {G}_{{\sigma }'_2 
{\sigma }'_1 \sigma _1 \sigma _2 }^{(\ref{eq1})cc} (\omega )$ and $\tilde 
{G}_{{\sigma }'\sigma }^{(2)cc} (\omega )$ finally we have
\sat
\begin{align}
& \tilde {G}_{{\sigma }'\sigma }^{(0)cc} (\omega ) = \frac{\delta _{{\sigma 
}'\sigma } }{\omega - E - \Sigma (\omega + io)}\,, \\ 
&\tilde {G}_{{\sigma }'_2 {\sigma }'_1 \sigma _1 \sigma _2 }^{(\ref{eq1})cc} (\omega 
) = \frac{\delta _{ - \sigma _1 \sigma _2 } \delta _{ - {\sigma }'_1 {\sigma 
}'_2 } (\delta _{{\sigma }'_1 \sigma _1 } - \delta _{{\sigma }'_1 \sigma _2 
} )}{\omega - E - U - \Sigma (\omega + io)} + \frac{\delta _{{\sigma }'_1 
\sigma _2 } \delta _{{\sigma }'_2 \sigma _1 } }{\omega - E - \Sigma (\omega 
- io)}\,, \\ 
&\tilde {G}_{{\sigma }'\sigma }^{(2)cc} (\omega ) = \frac{\delta _{{\sigma 
}'\sigma } }{\omega - E - U - \Sigma (\omega - io)}\, \cdot
 \end{align}

\vs
\ce{\textbf{VI. CONCLUSION}}

\vs
The expressions (103)\,-\,(105) of the (generalized) Green functions with the 
contributions of the high order approximations are similar to the expression 
(25)\,-\,(27) of these functions in the lowest order of the perturbation 
theory. The appearance of the self-energy part $\Sigma (\omega )$ in the 
denominators of the (generalized) Green functions is the consequence of the 
tunneling of electrons between the dot and the leads which takes place in 
high order approximations. Instead of the expression (38) of the function 
$W$(\textit{$\omega $}) determining the transport current now we have 
\begin{align}
W(\omega ) &= \sum\limits_{\pmb k} {2\pi \,\delta \,[\omega - E_a 
({\pmb k})]\;\left| {V_a ({\pmb k})} \right|^2} \,\sum\limits_{{\pmb k}'} 
{2\pi \,\delta \,[\omega - E_b ({{\pmb k}'})]\;\left| {V_b ({{\pmb k}'})} 
\right|^2} \, \cdot\nn \\
&\quad \frac{1}{Z_c }\left\{ {\left| {\frac{1}{\omega - E - \Sigma (\omega + 
io)\,}} \right|^2 + } \right.e^{ - \,\beta {E}'}\,\left( {\left| 
{\frac{1}{\omega - E - U - \Sigma (\omega + io)\,}} \right|^2} \right. \nn \\ 
& + \left| {\frac{1}{\omega - E - \Sigma (\omega - io)\,}} 
\right|^2 + \left. {\left| {\frac{1}{\omega - E - U - \Sigma (\omega + io)} 
- \frac{1}{\omega - E - \Sigma (\omega - io)\;}} \right|^2} \right)\; \nn \\ 
& + e^{ - \,\beta (2{E}' + U)}\,\left. {\left| {\frac{1}{\omega - 
E - U - \Sigma (\omega - io)\;}} \right|^2} \right\} \, \cdot 
\end{align}

In comparison with other methods the S-matrix approach presented in this 
work has several advantages. The strong Coulomb interaction between the 
electrons in the dot was taken into account exactly without the use of the 
perturbation theory. All crossing terms describing the co-tunneling are 
included automatically so that there is no need to assume the non-crossing 
approximation (NCA). We have derived the expression of the electrical 
current generated by the single-electron tunneling between the dot and the 
leads. The role of the multi-electron co-tunneling will be studied in a 
subsequent work. Moreover, it is easy to extend the presented method to the 
study of the multi-level quantum dots as well as to other nanosystems. 

Since the electron spin projections are explicitly indicated in the 
expressions of the matrix elements, it is convenient to apply the presented 
method to the study of the spin-dependent transition (transport) processes 
in the spintronic materials and devices. By including the interactions of 
the electrons with the electromagnetic field it is straightforward to extend 
the presented method to the study of the photon-assisted transition 
(transport) processes and develop the S-matrix approach in the theory of 
nanophotonic materials and devices.

\vss
\n \textbf{{Acknowledgment.}} The authors would like to express their thanks to the National Natural Sciences Council of Vietnam for the support to this work.

\end{document}